\documentclass[10pt]{iopart}
\usepackage{amsfonts}
\usepackage{amssymb}
\usepackage{graphicx}

\begin{document}

\title{Second-order light deflection by tidal charged black holes on the
brane}
\author{L\'{a}szl\'{o} \'{A}rp\'{a}d Gergely$^{1,2\star }$, Zolt\'{a}n
Keresztes$^{1,2\P }$, Marek Dwornik$^{1,2\dag }$}

\address{
$^{1}$ Department of Theoretical Physics, University of Szeged, Tisza Lajos krt 84-86, Szeged 6720, Hungary\\
$^{2}$ Department of Experimental Physics, University of Szeged, D\'{o}m t\'{e}r 9, Szeged 6720, Hungary\\
$^{\star }$ gergely@physx.u-szeged.hu, 
$^{\P }$zkeresztes@titan.physx.u-szeged.hu, 
$^{\dag }~$marek@titan.physx.u-szeged.hu}

\begin{abstract}
We derive the deflection angle of light rays passing near a black hole with
mass $m$ and tidal charge $q$, confined to a generalized Randall-Sundrum
brane with codimension one. We employ the weak lensing approach, up to the
second order in perturbation theory by two distinct methods. First we adopt
a Lagrangian approach and derive the deflection angle from the analysis of
the geodesic equations. Then we adopt a Hamiltonian approach and we recover
the same result from the analysis of the eikonal. With this we re-establish
the unicity of the result as given by the two methods. Our results in turn
implies a more rigurous constraint on the tidal charge from Solar System
measurements, then derived before.
\end{abstract}

\maketitle

\section{Introduction}

The possibility that the gravitational interaction acts in a more than
four-dimensional non-compact space-time \cite{RS}, while keeping the other
interactions locked in four space-time dimensions, has raised interesting
new perspectives in the solvability of the hierarchy problem and in
cosmological evolution \cite{BDEL}. This hypothesis has led to alternative
explanations for dark matter \cite{MakHarko}, \cite{Kar} and \cite{Pal}.

The most common such brane-world model is five-dimensional, containing a
four-dimensional (time-evolving three-dimensional) brane. In the context of
such generalized Randall-Sundrum brane-worlds, both the five-dimensional
space-time and the embedded brane are allowed to have generic curvature.
Gravitational dynamics on the brane is governed by an effective Einstein
equation \cite{SMS}, derived in full generality in \cite{Decomp}, \cite%
{VarBraneTensionPRD}. Higher codimension branes were also considered in
connection with conical singularities \cite{BGNS}, \cite{deRham}.

Spherically symmetric brane-world black holes were studied both by
analytical and numerical methods. Six-dimensional vacuum black holes, which
are locally higher dimensional Schwarzschild, were found in Ref. \cite%
{Kaloper}. A shell-like distribution of five-dimensional matter (violating
both the weak and strong energy conditions in the vicinity of the brane, but
falling off at infinity) is able to support a static five-dimensional black
hole localized on the brane, with the horizon rapidly "decaying" in the
extra dimension, \cite{KantiTamvakis}, \cite{KantiOlasagastiTamvakis}. Black
holes with a radiating component in the extra dimension were investigated in
connection with the AdS/CFT correspondence \cite{EmparanFabbriKaloper}. A
numerical analysis in an asymptotically five-dimensional Anti de Sitter
space-time showed the existence of small brane black holes (compared to the
five-dimensional curvature) as five-dimensional Schwarzschild solutions \cite%
{KudohTanakaNakamura}, however at large masses the effect of the brane
forbids to find such solutions. The generic static, spherically symmetric,
five-dimensional vacuum black hole on a three-dimensional time-evolving
brane, as a solution of the full set of the five-dimensional Einstein
equations was not found yet.

A static, spherically symmetric, vacuum black hole solution of the effective
Einstein equation\footnote{%
Besides having no matter on the brane, any five-dimensional matter source
could be present only if the pull-back of its energy-momentum tensor to the
brane vanishes. Also, the embedding of the brane is symmetric.} has been
found \cite{tidalRN}, but the five-dimensional space-time in which it can be
embedded, remains unknown. Beside the mass $m$, this black hole has a 
\textit{tidal }charge $q$ arising from the Weyl curvature of the higher
dimensional space-time, as shown in the line element 
\begin{equation}
ds^{2}=-f\left( r\right) dt^{2}+f^{-1}\left( r\right) dr^{2}+r^{2}\left(
d\theta ^{2}+\sin ^{2}\theta d\varphi ^{2}\right) \,~,  \label{RN}
\end{equation}%
with the metric function 
\begin{equation}
f\left( r\right) =1-\frac{2m}{r}+\frac{q}{r^{2}}\,.
\end{equation}

Formally the metric (\ref{RN}) is the Reissner-Nordstr\"{o}m solution of a
spherically symmetric Einstein-Maxwell system in general relativity. There,
however the place of the tidal charge $q$ is taken by the square of the 
\textit{electric} charge $Q$. Thus $q=Q^{2}$ is always positive, when the
metric (\ref{RN}) describes the spherically symmetric exterior of an
electrically charged object in general relativity. By contrast, in
brane-world theories the metric (\ref{RN}) allows for any $q$.

When $\left\vert q\right\vert \gg 2mr$ (thus for a light black hole and not
too far from its horizon) and $q<0$ the tidal charge term dominates. This
remark is consistent with the result of Ref. \cite{KudohTanakaNakamura},
according to which five-dimensional Schwarzschild solutions (with $%
f=1-m_{5}/r^{2}$) can be considered as brane black holes as long as their
mass is small compared to the five-dimensional curvature radius.

On the other hand when $\left\vert q\right\vert \ll 2mr$, the term
containing the tidal charge becomes a mere correction of the
four-dimensional Schwarzschild metric. The four-dimensional Schwarzschild
metric can be extended into the fifth dimension as a black string \cite{ChRH}%
, which solves the five-dimensional Einstein equations. Gravity wave
perturbations of such a black-string brane-world were computed in Ref. \cite%
{SCMlet}. Due to the Gregory-Laflamme instability \cite{GL} the string is
expected to pinch off, thus a black cigar emerges \cite{Gregory} (although
under very mild assumptions, classical event horizons will not pinch off 
\cite{HorowitzMaeda}).

For $\left\vert q\right\vert \ll 2mr$ (small tidal charge and / or far from
the horizon) the correction to the Schwarzschild potential represented by
the tidal charge term scales with $r^{-2}$. On the other hand the
perturbative analysis of the gravitational field of a spherically symmetric
source in the weak field limit in the original Randall-Sundrum setup
(Schwarzschild black hole on a brane embedded in Anti de Sitter
five-dimensional space-time) has shown corrections to the Schwarzschild
potential scaling as $r^{-3}$ \cite{RS}, \cite{GarrigaTanaka} and \cite%
{Giddings}. Therefore the five-dimensional extension of the tidal charged
black hole is not a linearly perturbed five-dimensional Anti de Sitter
space-time \cite{BlackString}, but should rather be considered as a
perturbation of the black string / black cigar metric.

The case $q>0$ is in full analogy with the general relativistic
Reissner-Nordstr\"{o}m solution. For $q<m^{2}$ it describes tidal charged
black holes with two horizons at $r_{h}=m\pm \sqrt{\left( m^{2}-q\right) }$,
both below the Schwarzschild radius. For $q=m^{2}$ the two horizons coincide
at $r_{h}=m$ (this is the analogue of the extremal Reissner-Nordstr\"{o}m
black hole). In these cases it is evident that the gravitational deflection
of light and gravitational lensing is decreased by $q$. Finally there is a
new possibility forbidden in general relativity due to physical
considerations on the smallness of the electric charge. This is $q>m^{2}$
for which the metric (\ref{RN}) describes a naked singularity. Such a
situation can arise whenever the mass $m$ of the brane object is of small
enough, compared to the effect of the bulk black hole generating Weyl
curvature, and as such, tidal charge. Due to its nature, the tidal charge$\
q $ should be a more or less global property of the brane, which can contain
many black holes of mass $m\geq \sqrt{q}$ and several naked singularities
with mass $m<\sqrt{q}$.

For any $q<0$ there is only one horizon, at $r_{h}=m+\sqrt{\left(
m^{2}+\left\vert q\right\vert \right) }$. For these black holes, gravity is
increased on the brane by the presence of the tidal charge \cite{tidalRN}.
Light deflection and gravitational lensing are stronger than for the
Schwarzschild solution.

The metric (\ref{RN}) can be also considered as the exterior of a star. In
this case one does not have to worry about the existence or location of
horizons, as they would lie inside the star, where some interior solution
should be matched to the metric (\ref{RN}). The generic feature that a
positive (negative) tidal charge is weakening (strengthening) gravitation on
the brane, is kept.

Tidal charged brane black hole metrics were studied before as vacuum
exteriors for interior stellar solution \cite{GermaniMaartens}, with the
purpose of confrontation with solar system tests \cite{BoehmerHarkoLobo},
evolution of thin accretion disks in this geometry \cite{accretion} and in a
thermodynamical context \cite{tidalThermo}.

Gravitational lensing could provide a test of such brane-world solutions, in
particular it may turn useful in the study of the tidal charged black hole.
The determination of the tidal charge can indicate ways to extend this
solution into the fifth dimension. Both weak \cite{KarSinha}, \cite{MM} and
strong \cite{Whisker} gravitational lensing of various five-dimensional
black holes were discussed, the topic being reviewed in Ref. \cite%
{MajumdarMukherjee}.

In this paper we derive the deflection angle of light rays caused by brane
black holes with tidal charge (\ref{RN}). Generalizing previous approaches 
\cite{KarSinha}, \cite{MM}, we carry on this computation up to the second
order in the weak lensing parameters. As the metric (\ref{RN}) is static, we
consider only the second order gravielectric contributions, but no
gravimagnetic contributions, which are of the same order and would appear
due to the movement of the brane black holes. Gravimagnetic effects in the
general relativistic approach were considered in \cite{SchaferBartelmann}.

In Section 2 we present a Lagrangian approach, based on Ref. \cite{GeDa}. We
conclude this section with the derivation of the light deflection angle to
second order accuracy in both $m$ and $q$. The first order contributions are
in agreement with the results of Ref. \cite{Hobill}. The second order
contributions however differ from the corresponding result of Ref. \cite%
{BoehmerHarkoLobo}, obtained by a Hamilton-Jacobi approach. In order to sort
out this discrepancy, in Section 3 we carefully employ the eikonal method to
the required order. As consequence our previous Lagrangian result is
reproduced, re-establishing the unicity of the expression for light
deflection. Starting from the improved result, we could strengthen the
constraint on the brane tension in Section 4. Section 5 contains the
concluding remarks.

\section{Lagrangian approach}

\subsection{Light propagation}

Light follows null geodesics of the metric (\ref{RN}). Its equations of
motion can be derived either from the geodesic equations, or from the
Lagrangian given by $2\mathcal{L}=\left( ds^{2}/d\lambda ^{2}\right) $,
where $\lambda $ is a parameter of the null geodesic curve (see Chapter 3 of
Ref. \cite{Straumann}. Due to spherical and reflectional symmetry across the
equatorial plane, $\theta =\pi /2$ can be chosen. Thus 
\begin{equation}
0=2\mathcal{L}=-f\left( r\right) \dot{t}^{2}+f^{-1}\left( r\right) \dot{r}%
^{2}+r^{2}\dot{\varphi}^{2}\,.  \label{Lag}
\end{equation}%
(A dot represents derivative with respect to $\lambda $.) The cyclic
variables $t$ and $\varphi $ lead to the constants of motion $E$ and $L$ 
\begin{equation}
E\equiv -p_{t}=f\dot{t}\,,\qquad L\equiv p_{\varphi }=r^{2}\dot{\varphi}\,.
\label{first_integrals}
\end{equation}%
By inserting these into Eq. (\ref{Lag}), passing to the new radial variable $%
u=1/r$ and introducing $\varphi $ as a dependent variable, we obtain 
\begin{equation}
\left( u^{\prime }\right) ^{2}=\frac{E^{2}}{L^{2}}-u^{2}f\,\left( u\right)
\,,  \label{radial}
\end{equation}%
where a prime refers to differentiation with respect to $\varphi $. $%
\allowbreak $

Unless $u^{\prime }=0$ (representing a circular photon orbit),
differentiation of Eq. (\ref{radial}) gives 
\begin{equation}
u^{\prime \prime }=-uf\,-\frac{u^{2}}{2}\frac{df}{du}\,\,,  \label{radial2}
\end{equation}%
For $f=1$, when there is no gravitation at all (the metric (\ref{RN})
becomes flat), the above equation simplifies to $u^{\prime \prime }+u=0$,
which is solved for $u=u_{0}=b^{-1}\cos \varphi $. The impact parameter $b$
represents the closest approach of the star on the straight line orbit
obtained by disregarding the gravitational impact of the star (this is the
viewpoint an asymptotic observer will take, as the metric (\ref{RN}) is
asymptotically flat). The polar angle $\varphi $ is measured from the line
pointing from the centre of the star towards the point of closest approach.
With$\ u^{\prime }=0$ at the point of closest approach, given in the
asymptotic limit by $u=b^{-1}$, Eq. (\ref{radial}) with $m=0=q$ gives $b=L/E$%
.

\subsection{Perturbative solution}

Eq. (\ref{radial2}), written in detail, gives 
\begin{equation}
u^{\prime \prime }+u=3mu^{2}-2qu^{3}\,\,.  \label{lensing}
\end{equation}%
For studying weak lensing, we look for a perturbative solution in series of
the small parameters 
\begin{equation}
\varepsilon =mb^{-1}\,,\qquad \eta =qb^{-2}  \label{params}
\end{equation}%
in the form 
\begin{equation}
u=b^{-1}\cos \varphi +\varepsilon u_{1}+\eta v_{1}+\varepsilon
^{2}u_{2}+\eta ^{2}v_{2}+\varepsilon \eta w_{2}+\mathcal{O}\left(
\varepsilon ^{3},\eta ^{3},\varepsilon \eta ^{2},\varepsilon ^{2}\eta
\right) \,.  \label{ansatz}
\end{equation}%
The index on the unknown functions $u_{1},\,u_{2},\,v_{1},\,v_{2}$ and $%
w_{2} $ counts the perturbative order in which they appear. By inserting Eq.
(\ref{ansatz}) into the weak lensing equation (\ref{lensing}) we obtain the
relevant differential equations for the unknown functions. Up to the second
order in both small parameters these are: 
\begin{eqnarray}
\varepsilon &:&\qquad u_{1}^{\prime \prime }+u_{1}=3b^{-1}\cos ^{2}\varphi
\,,  \label{ep} \\
\eta &:&\qquad v_{1}^{\prime \prime }+v_{1}=-2b^{-1}\cos ^{3}\varphi \,,
\label{et} \\
\varepsilon ^{2} &:&\qquad u_{2}^{\prime \prime }+u_{2}=3u_{1}\left[
u_{1}\left( m-2qb^{-1}\cos \varphi \right) +2\cos \varphi \right] \,,
\label{ep2} \\
\eta ^{2} &:&\qquad v_{2}^{\prime \prime }+v_{2}=3v_{1}\left[ v_{1}\left(
m-2qb^{-1}\cos \varphi \right) -2\cos ^{2}\varphi \right] \,,  \label{et2} \\
\varepsilon \eta &:&\qquad w_{2}^{\prime \prime }+w_{2}=6\bigl[%
u_{1}v_{1}\left( m-2qb^{-1}\cos \varphi \right)  \nonumber \\
&&\qquad +v_{1}\cos \varphi -u_{1}\cos ^{2}\varphi \bigr]\,.  \label{epet}
\end{eqnarray}%
$\allowbreak $Note that the solutions are not allowed to contain
contributions with the property $f\left( -\varphi \right) =-f\left( \varphi
\right) $, as the zero of $\varphi $ was chosen at the point of closest
approach $r_{\min }$, with respect to which the past and future portions of
the path are symmetric (this is a consequence of the static nature of the
lensing metric (\ref{RN}).) The first order equations (\ref{ep}) and (\ref%
{et}) are solved for$\allowbreak $ 
\begin{eqnarray}
u_{1} &=&b^{-1}\left[ C_{\varepsilon }\cos \varphi +\frac{1}{2}\left( 3-\cos
2\varphi \right) \right] \,,  \label{u1} \\
v_{1} &=&b^{-1}\left[ C_{\eta }\cos \varphi +\frac{1}{16}\left( \cos
3\varphi -12\varphi \sin \varphi \right) \right] \,,  \label{v1}
\end{eqnarray}%
where $C_{\varepsilon ,~\eta }$ are constants of integration appearing at
the order shown by their indices and we have dropped (by choosing as zero
their pre-factors, which are constants of integration) the terms
proportional to $\sin \varphi $, in accordance with the earlier remark. To
keep all terms in $u_{1}$ and $v_{1}$ of comparable order, we have factored
out $b^{-1}$ from the constants. Thus, both $mu_{1}$ and $mv_{1}$ are of
order $\varepsilon $, while both $qb^{-1}u_{1}$ and $qb^{-1}v_{1}$ are of
order $\eta $. In consequence, all these terms drop out from Eqs. (\ref{ep2}%
)-(\ref{epet}), which are then solved for 
\begin{eqnarray}
u_{2} &=&b^{-1}\left[ C_{\varepsilon ^{2}}\cos \varphi +C_{\varepsilon
}\left( 3-\cos 2\varphi \right) +\frac{3}{16}\left( \cos 3\varphi +20\varphi
\sin \varphi \right) \right] \,,  \label{u2} \\
v_{2} &=&b^{-1}\bigl[C_{\eta ^{2}}\cos \varphi +\frac{3}{16}C_{\eta }\left(
\cos 3\varphi -12\varphi \sin \varphi \right)  \nonumber \\
&&+\frac{1}{256}\bigl(-21\cos 3\varphi +\cos 5\varphi +60\varphi \sin \varphi
\nonumber \\
&&-36\varphi \sin 3\varphi -72\varphi ^{2}\cos \varphi \bigr)\bigr]\,,
\label{v2} \\
w_{2} &=&b^{-1}\bigl[C_{\varepsilon \eta }\cos \varphi +C_{\eta }\left(
3-\cos 2\varphi \right) +\frac{3}{16}C_{\varepsilon }\left( \cos 3\varphi
-12\varphi \sin \varphi \right)  \nonumber \\
&&+\frac{1}{16}\left( -60+31\cos 2\varphi -\cos 4\varphi +12\varphi \sin
2\varphi \right) \bigr]\,,  \label{w2}
\end{eqnarray}%
where $C_{\varepsilon ^{2},~\eta ^{2},~\varepsilon \eta }$ represent
additional constants of integration (and as before, we have dropped $\sin
\varphi $ terms, which also arise by integration). The remaining integration
constants can be fixed by inserting the solution (\ref{ansatz}) with the
coefficients (\ref{u1})-(\ref{w2}) into Eq. (\ref{radial}). They are $%
\allowbreak \allowbreak \allowbreak C_{\varepsilon }=C_{\varepsilon \eta
}=0,~C_{\eta }=-9/16,~C_{\varepsilon ^{2}}=37/16,~C_{\eta ^{2}}=271/256$.
With this, we have found the generic solution of Eq. (\ref{radial}), up to
the second order in both small parameters:

\begin{eqnarray}
bu &=&\cos \varphi +\frac{\varepsilon }{2}\left( 3-\cos 2\varphi \right) -%
\frac{\eta }{16}\left( 9\cos \varphi -\cos 3\varphi +12\varphi \sin \varphi
\right)  \nonumber \\
&&+\frac{\varepsilon ^{2}}{16}\left( 37\cos \varphi +3\cos 3\varphi
+60\varphi \sin \varphi \right)  \nonumber \\
&&+\frac{\eta ^{2}}{256}\bigl(271\cos \varphi -48\cos 3\varphi +\cos 5\varphi
\nonumber \\
&&+384\varphi \sin \varphi -36\varphi \sin 3\varphi -72\varphi ^{2}\cos
\varphi \bigr)  \nonumber \\
&&+\frac{\varepsilon \eta }{16}\left( -87+40\cos 2\varphi -\cos 4\varphi
+12\varphi \sin 2\varphi \right) ~.  \label{u}
\end{eqnarray}%
(Note that the coefficients of $\cos \varphi $ in the $\varepsilon ^{2}$ and 
$\eta ^{2}$ terms are corrected with respect to reference \cite{GeDa}, where
in the solution of Eq. (\ref{lensing}) the choice of the constants $%
C_{\varepsilon ^{2}}$ and $C_{\eta ^{2}}$ was not verified to solve Eq. (\ref%
{radial}).)

$\allowbreak $Far away from the lensing object $u=0$ and $\,\varphi =\pm \pi
/2\pm \delta \varphi /2$, where the $+\left( -\right) $ sign is for the
light signal in the distant future (past), and $\delta \varphi $ represents
the angle with which the light ray is bent by the lensing object with mass $%
m $ and tidal charge $q$. In our second-order approach this has the form:\ 
\begin{equation}
\delta \varphi =\varepsilon \alpha _{1}+\eta \beta _{1}+\varepsilon
^{2}\alpha _{2}+\eta ^{2}\beta _{2}+\varepsilon \eta \gamma _{2}+\mathcal{O}%
\left( \varepsilon ^{3},\eta ^{3},\varepsilon \eta ^{2},\varepsilon ^{2}\eta
\right) \,.
\end{equation}%
A power series expansion of the solution (\ref{u}) in the small parameters
then gives the coefficients of the above expansion, and the deflection angle
becomes:%
\begin{equation}
\delta \varphi =4\varepsilon -\frac{3\pi }{4}\eta +\frac{15\pi }{4}%
\varepsilon ^{2}+\frac{105\pi }{64}\eta ^{2}-16\varepsilon \eta \,.
\label{phivariation}
\end{equation}%
The first three terms of this expansion were already given in \cite{Hobill}
for the Reissner-Nordstr\"{o}m black hole. There, however the argument that $%
\eta $ is of $\epsilon ^{2}$ order was advanced. In brane-worlds there is no
a priori reason for considering only small values of the tidal charge, thus
we have computed the deflection angle $\delta \varphi $ containing all
possible contributions up to second order in both parameters.

The deflection angle however is given in terms of the Minkowskian impact
parameter $b$. It would be useful to write this in term of the distance of
minimal approach $r_{\min }$ as well. The minimal approach is found by
inserting the values $u=1/r_{\min }$ and $\varphi =0$ in Eq. (\ref{u}):%
\begin{equation}
r_{\min }=b\left( 1\allowbreak -\varepsilon +\allowbreak \frac{1}{2}\eta -%
\frac{3}{2}\varepsilon ^{2}-\frac{5}{8}\eta ^{2}+2\varepsilon \eta \right) ~.
\end{equation}%
Inverting this formula gives to second order accuracy (the small parameters
being now $m/r_{\min }$ and $q/r_{\min }^{2}$):%
\begin{equation}
\frac{1}{b}=\frac{1}{r_{\min }}\left( 1\allowbreak -\frac{m}{r_{\min }}%
\allowbreak +\allowbreak \frac{q}{2r_{\min }^{2}}\allowbreak -\frac{m^{2}}{%
2r_{\min }^{2}}-\frac{q^{2}}{8r_{\min }^{4}}+\frac{mq}{2r_{\min }^{3}}%
\right) ~.
\end{equation}%
As the deflection angle consists only of first and second order
contributions, the above formula is needed only to first order for
expressing $\delta \varphi $ in terms of the minimal approach: 
\begin{equation}
\delta \varphi =\allowbreak \allowbreak \frac{4m}{r_{\min }}-\frac{3\pi q}{%
4r_{\min }^{2}}\allowbreak \allowbreak +\frac{\left( 15\pi -16\right) m^{2}}{%
4r_{\min }^{2}}+\frac{57\pi q^{2}}{64r_{\min }^{4}}+\frac{\left( 3\pi
-28\right) mq}{2r_{\min }^{3}}\,.  \label{deflection2}
\end{equation}%
The first three terms again agree with the ones given in \cite{Hobill}, when 
$q=Q^{2}$.

\section{Hamiltonian approach}

In this section we employ the eikonal method for deriving the light
deflection angle. Subsection 3.1 follows the derivation presented in Ref. 
\cite{BoehmerHarkoLobo}, which in turn is based on Ref. \cite{Landau}.
Instead of the coordinate transformation employed in Ref. \cite%
{BoehmerHarkoLobo}, in subsection 3.2 we follow a perturbative approach
based on a double expansion in both small parameters. This ensures a higher
accuracy of the perturbative result, and we recover the deflection angle (%
\ref{deflection2}). Some of the technical details are presented in the
Appendix.

\subsection{Light deflection from the eikonal equation}

This derivation starts from the general relativistic eikonal equation
(Hamilton-Jacobi equation)%
\begin{equation}
g^{ab}\frac{\partial \Psi }{\partial x^{a}}\frac{\partial \Psi }{\partial
x^{b}}=0~.  \label{eikonal}
\end{equation}%
Here the function $\Psi $ is the rapidly varying real phase of the complex
electromagnetic 4-potential $A_{a}=$\textrm{Re}$\left[ \mathcal{A}_{a}\exp
\left( i\Psi \right) \right] $ and $k_{a}=\partial \Psi /\partial x^{a}$ is
the wave vector. The complex amplitude varies only slowly in the geometrical
optics (eikonal / high-frequency) approximation: the wavelength is small
compared to either of the characteristic curvature radius or the typical
length of variation of the optical properties. The normalized version of the
complex amplitude is the polarization vector. Light rays are defined as the
integral curves of $k^{a}$ and they are perpendicular to the wave-fronts,
the surfaces of constant phase. The eikonal equation then is but the
condition that the wave vector is null. The vacuum Maxwell equations further
imply that light travels along null geodesics, $k^{a}\nabla _{a}k^{b}=0$ and
the polarization vector is perpendicular to the light rays and parallel
propagated along them (fore more details see Chapter 1.8. of Ref. \cite%
{Straumann}.

As before, we discuss orbits in the equatorial plane\ $\theta =\pi /2$. Due
to the symmetries of the problem the eikonal function (which is the
Hamilton-Jacobi action) can be chosen as 
\begin{equation}
\Psi =E\left( -t+b\varphi \right) +\psi _{r}\left( r\right) ~,  \label{Psi1}
\end{equation}%
where we have employed the relation $L=bE$ derived at the end of subsection
2.1.

The eikonal equation (\ref{eikonal}) gives for the unknown radial function 
\begin{eqnarray}
\psi _{r} &=&E\int \mathcal{C}\left( r\right) dr~, \\
\mathcal{C}\left( r\right) &=&\pm \sqrt{\frac{r^{4}}{\left(
r^{2}-2b\varepsilon r+b^{2}\eta \right) ^{2}}-\frac{b^{2}}{%
r^{2}-2b\varepsilon r+b^{2}\eta }}~.  \label{psir}
\end{eqnarray}%
Here we have employed the definition (\ref{params}) of the small parameters $%
\varepsilon $ and $\eta $ and we choose the negative root in front of the
square root when $r$ decreases (the photon approaches the lensing object)
and the positive root when $r$ increases (the photon has already overpassed
the lensing object). This choice assures $d\psi _{r}\left( r\right) =E%
\mathcal{C}\left( r\right) dr>0$, regardless whether the photon is
approaching or departing from the point of closest approach.

By differentiating Eq. (\ref{Psi1}) with respect to $L$ we obtain%
\begin{equation}
\frac{d\Psi }{dL}=\varphi +\frac{d\psi _{r}}{dL}~,  \label{PsiL}
\end{equation}%
however due to Jacobi's Theorem the derivative of the Hamilton-Jacobi action
with respect to a canonical constant (in this case $L$), gives another
canonical constant. [$\Psi $ is the generating function of the canonical
transformation to pairs of variables trivially obeying the Hamiltonian
equations; in the present case $L$ and $d\Psi /dL$ such that $\left(
d/d\lambda \right) L=0=\left( d/d\lambda \right) \left( d\Psi /dL\right) $,
where $\lambda $\thinspace\ is a parameter along the trajectory of the
photon.]

For later convenience we introduce a new auxiliary variable $\phi $ by $%
r=b/\cos \phi $. To zeroth order in the small parameters the relation $%
\varphi =\phi $ holds, see the remarks at the end of subsection 2.1. As $%
r>0\,$\ we have $\arccos \left( b/r\right) =\left\vert \phi \right\vert $,
in other words $\phi =~\mathrm{sgn}\phi \arccos \left( b/r\right) $. The
variable $\phi $ exists for any $r\geq b$. By evaluating Eq. (\ref{PsiL}) at
two points $r\geq b$ on the trajectory $\varphi \left( \phi \right) $ and
forming the difference leads to%
\begin{equation}
\varphi \left( \phi _{2}\right) -\varphi \left( \phi _{1}\right) =-\left. 
\frac{d\psi _{r}}{dL}\right\vert _{\phi _{2}}+\left. \frac{d\psi _{r}}{dL}%
\right\vert _{\phi _{1}}~.  \label{PsiL1}
\end{equation}

While the photon travels from the infinity to the point of closest approach $%
r=r_{\min }$ and then back to infinity, the total change in the polar angle $%
\varphi $ can be found in a limiting process as%
\begin{equation}
\Delta \varphi =-\lim_{\Phi \rightarrow \pi /2}\left( \left. \frac{\partial
\psi _{r}}{\partial L}\right\vert _{\Phi }-\left. \frac{\partial \psi _{r}}{%
\partial L}\right\vert _{-\Phi }\right) ~.  \label{DeltaPhi}
\end{equation}%
(Here $\Phi \geq 0$.)

\subsection{Perturbative solution}

We expand the radial function $\psi _{r}$ given by (\ref{psir})\ to second
order accuracy in both $\varepsilon $ and $\eta $: 
\begin{eqnarray}
\psi _{r} &=&L\int \mathcal{C}\left( \phi \right) \frac{\sin \phi }{\cos
^{2}\phi }d\phi ~, \\
\mathcal{C}\left( \phi \right) &=&\sin \phi +\left( \varepsilon -\frac{1}{2}%
\eta \cos \phi \right) \frac{\left( 2-\cos ^{2}\phi \right) \cos \phi }{\sin
\phi }\allowbreak  \nonumber \\
&&+\frac{1}{2}\left( \varepsilon -\frac{1}{2}\eta \cos \phi \right) ^{2}%
\frac{\cos ^{2}\phi }{\sin ^{3}\phi }  \nonumber \\
&&\times \left[ 4\left( 3-\cos ^{2}\phi \right) \sin ^{2}\phi -\left( 2-\cos
^{2}\phi \right) ^{2}\right] \allowbreak ~.  \label{psi_phi_integrand}
\end{eqnarray}%
The expression $\mathcal{C}\left( \phi \right) \equiv \mathcal{C}\left(
r=b/\cos \phi \right) $ changes sign with $\mathrm{sgn}\phi $, in accordance
with the assumption made in Eq. (\ref{psir}) for the sign in front of the
square root.

By the computation presented in the Appendix it is immediate to derive the
zeroth, $\varepsilon ,~\eta ,~\varepsilon ^{2},~\eta ^{2},~\varepsilon \eta $
order contributions to $\Delta \varphi $ as $\left( \pi ,~\!4,~\!\!-\!3\pi
/4,~\!15\pi /4,~\!105\pi /64,~\!\!-\!16\right) $.

To zeroth order we have found that $\left( \Delta \varphi \right) _{0}=\pi $%
, thus the path of the photon in the absence of the perturbing object with
mass $m$ and tidal charge $q$ is a straight line. Thus the deflection caused
by the\textbf{\ }mass and tidal charge of the lensing object\ when the
photon travels from the infinity to the nearest point $r=r_{\min }$ and then
back to the infinity is given by\textbf{\ }$\delta \varphi =\Delta \varphi
-\pi $\textbf{.}

In consequence the term-by-term computation by the limiting process (\ref%
{DeltaPhi}) reproduces exactly the deflection angle (\ref{phivariation}),
derived earlier in a Lagrangian approach.

\section{Solar system constraints}

The most important difference in comparing Eq. (\ref{phivariation}) to Eq.
(27) of Ref. \cite{BoehmerHarkoLobo} is the presence of the $\eta $-term in
our result, which turns out to be the dominant contribution to the
deflection angle caused by the tidal charge. This is $\varepsilon ^{-1}$
times larger than the $\varepsilon \eta $ mixed term.

This implies that the constraints on the tidal charge derived in Ref. \cite%
{BoehmerHarkoLobo}, as imposed by the measurements of the deflection of
light by the Sun should be re-evaluated. For this we follow the logic of Ref 
\cite{BoehmerHarkoLobo}, but apply the observational constraint to the $\eta 
$ term (rather than $\varepsilon \eta $). Long baseline radio interferometry
measurements \cite{RadioInt}-\cite{RadioInt2} give $\delta \varphi =\delta
_{\varepsilon }\varphi \left( 1+\xi \right) $, with $\xi <$ $\xi _{\max
}=\pm 0.0017$.\textbf{\ }By assuming that the dominant deviation from the
(first order) Schwarzschild value is due to the tidal charge, we obtain: $%
\delta _{\varepsilon }\varphi \xi _{\max }=\left( \delta _{\eta }\varphi
\right) _{\max }$, thus $16\varepsilon \xi _{\max }=3\pi \left( -\eta
\right) _{\max }$ or $16mb\xi _{\max }=3\pi \left( -q\right) _{\max }$. With
the mass of the Sun $m=M_{\odot }=1476.685\mathrm{m}$ and the smallest
possible impact parameter (equal to the closest possible approach $r_{\min
}=R_{\odot }=$ $695990$ $\mathrm{km}$), in the first order approximation
employed here:%
\begin{equation}
\left\vert q\right\vert _{\max }=\frac{16\left\vert \xi _{\max }\right\vert 
}{3\pi }M_{\odot }R_{\odot }=2966\mathrm{km}^{2}~.  \label{qmax}
\end{equation}%
The junction conditions of the tidal charged brane black hole with a star of
uniform density $\rho \,$ (which dominates over the tidal contribution to
the energy density), applied for the Sun give a negative tidal charge \cite%
{GermaniMaartens}:%
\begin{equation}
q_{\odot }=-\frac{3M_{\odot }R_{\odot }\rho _{\odot }}{\lambda }~.  \label{q}
\end{equation}%
From $-q_{\odot }\leq \left\vert q\right\vert _{\max }$ we find%
\begin{equation}
\lambda \geq \frac{3M_{\odot }R_{\odot }\rho _{\odot }}{\left\vert
q\right\vert _{\max }}=\frac{9\pi \rho _{\odot }}{16\left\vert \xi _{\max
}\right\vert }=1464.066\frac{\mathrm{g}}{\mathrm{cm}^{3}}=6.310\cdot 10^{-3}%
\mathrm{MeV}^{4}~.
\end{equation}%
The constraint on the brane tension from Solar System measurements is
therefore 5 orders of magnitude stronger than derived in Ref. \cite%
{BoehmerHarkoLobo}, however still far weaker than all other constraints ($%
\lambda \geq 138.59$ \textrm{TeV}$^{4}$ from table-top experiments \cite%
{braneworldgrav}, \cite{IrradFriedmann}, $\lambda \geq 1$\textrm{MeV}$^{4}$
from nucleosynthesis \cite{BBNLim} and $\lambda \geq 5\times 10^{8}$ \textrm{%
MeV}$^{4}$ from neutron stars \cite{GermaniMaartens}).

\section{Concluding remarks}

In this paper we have computed the light deflection angle due to a tidal
charged brane black hole /\ naked singularity / star (depending on $q$), up
to second order in the two small parameters $\varepsilon $ and $\eta $,
related to the mass and tidal charge of the lensing object. We have carried
on this task by two distinct methods and obtained identical results.

The first method relies on a Lagrangian, while the second on a
Hamilton-Jacobi approach. The latter was first applied in Ref. \cite%
{BoehmerHarkoLobo}, however that calculation focused only on the first order
correction in $q$ of the Schwarzschild deflection angle (thus an $%
\varepsilon \eta $-contribution in our terminology), given in their Eq. (27).

In comparison, besides the Schwarzschild contribution $\varepsilon $, our
result (\ref{phivariation}) for the light deflection angle contains the
second order Schwarzschild correction $\varepsilon ^{2}$, the first and
second order tidal contributions $\eta $ and $\eta ^{2}$, finally the mixed
contribution $\varepsilon \eta $. The latter turns out to be twice the value
given in Ref. \cite{BoehmerHarkoLobo}, where the expansion was not performed
everywhere to this order (for example in the Jacobian of the transformation
involved there). As a consequence of these improvements we have strengthened
the limit imposed on the brane tension by Solar System measurements by 5
orders of magnitude.

As already remarked in \cite{Sereno}, the electric charge of the
Reissner-Nordstr\"{o}m black hole decreases the deflection angle, as
compared to the Schwarz\-schild case. The same holds true for a \textit{%
positive} tidal charge. If the condition $16mr_{\min }=3\pi q$ is obeyed,
the first order contributions to the deflection angle cancel (there is no
deflection to first order) and the three second order terms of $\delta
\varphi $ presented in this paper give the leading effect to weak lensing.

Furthermore, $16mr_{\min }<3\pi q$ could be obeyed, leading to a \textit{%
negative} deflection angle (to first order). That would mean that rather
than magnifying distant light sources, such a lensing object will demagnify
them.

By contrast, a negative tidal charge can considerably increase the lensing
effect. Such negative tidal charged brane black holes arise naturally as
exteriors of static brane stars composed of ideal fluid with constant
density \cite{GermaniMaartens}. Therefore a negative tidal charge could be
responsible at least for part of the lensing effects attributed at present
to dark matter. However in line with the Solar System constraint (\ref{qmax}%
) imposed in this paper on the tidal charge of the Sun, we do not expect
these contributions to be very large.

\ack

This work was supported by the Hungarian OTKA grant No.~69036. L\'{A}G was
further supported the London South Bank University Research Opportunities
Fund and the Pol\'{a}nyi Program of the Hungarian National Office for
Research and Technology (NKTH).

\appendix

\section{The change in the polar angle}

The radial contribution (\ref{psi_phi_integrand}) to the eikonal function
can be written as the series 
\begin{equation}
\psi _{r}\left( \phi \right) =L\left( I_{0}+\varepsilon I_{\varepsilon
}+\eta I_{\eta }+\varepsilon ^{2}I_{\varepsilon ^{2}}+\eta ^{2}I_{\eta
^{2}}+\varepsilon \eta I_{\varepsilon \eta }\right) ~,
\end{equation}%
with$\allowbreak \allowbreak $%
\begin{eqnarray}
I_{0}\left( \phi \right)  &=&\tan \phi -\phi ~,  \nonumber \\
I_{\varepsilon }\left( \phi \right)  &=&\allowbreak \ln \left( \frac{1+\sin
\phi }{1-\sin \phi }\right) -\sin \phi   \nonumber \\
&=&2\mathrm{sgn}\phi ~\mathrm{arccosh}\left( \frac{1}{\cos \phi }\right)
-\sin \phi ~,  \nonumber \\
I_{\eta }\left( \phi \right)  &=&\frac{1}{4}\sin \phi \cos \phi -\frac{3\phi 
}{4}~,  \nonumber \\
I_{\varepsilon ^{2}}\left( \phi \right)  &=&\frac{15}{4}\phi +\left( 3\cos
^{2}\phi -1\right) \frac{\cot \phi }{4}~,  \nonumber \\
I_{\eta ^{2}}\left( \phi \right)  &=&\frac{35}{64}\phi +\left( 6\cos
^{4}\phi -33\cos ^{2}\phi +35\right) \frac{\cot \phi }{64}~,  \nonumber \\
I_{\varepsilon \eta }\left( \phi \right)  &=&\frac{8\cos ^{2}\phi -\cos
^{4}\phi -8}{2\sin \phi }~.  \label{Iet}
\end{eqnarray}%
As all of these terms are antisymmetric, the radial contribution to the
eikonal function is also antisymmetric, $\psi _{r}\left( -r\right) =-\psi
_{r}\left( r\right) $.

Besides the explicit global factor $L$, the expression $\psi _{r}\left( \phi
\right) $ also depends on $L$ through $\varepsilon =mb^{-1}$, $\eta =qb^{-2}$
and $\phi =\mathrm{sgn}\phi \arccos \left( b/r\right) $, as $b=L/E$. However
the products $L\varepsilon $ and $L^{2}\eta $ are independent of $L$. In
consequence the derivative $d\psi _{r}\left( \phi \right) /dL$ can be
calculated as%
\begin{eqnarray}
\frac{d}{dL}\psi _{r}\left( \phi \right) &=&\frac{d}{dL}\left( LI_{0}\right)
+\left( L\varepsilon \right) \frac{d}{dL}I_{\varepsilon }+\left( L^{2}\eta
\right) \frac{d}{dL}\left( L^{-1}I_{\eta }\right)  \nonumber \\
&&+\left( L\varepsilon \right) ^{2}\frac{d}{dL}\left( L^{-1}I_{\varepsilon
^{2}}\right) +\left( L^{2}\eta \right) ^{2}\frac{d}{dL}\left( L^{-3}I_{\eta
^{2}}\right)  \nonumber \\
&&+\left( L\varepsilon \right) \left( L^{2}\eta \right) \frac{d}{dL}\left(
L^{-2}I_{\varepsilon \eta }\right) ~.
\end{eqnarray}%
The term-by-term computation, by employing the identity 
\begin{equation}
\frac{d\phi }{dL}=-L^{-1}\cot \phi ~,
\end{equation}%
gives%
\begin{eqnarray}
\frac{d}{dL}\left( LI_{0}\right) &=&-\phi ~,  \nonumber \\
L\frac{d}{dL}I_{\varepsilon } &=&\frac{-2}{\sin \phi }+\frac{\cos ^{2}\phi }{%
\sin \phi }~,  \nonumber \\
L^{2}\frac{d}{dL}\left( L^{-1}I_{\eta }\right) &=&\frac{3\phi }{4}-\frac{%
\cos ^{2}\phi -3}{4\mathrm{\sin }\phi }\cos \phi ~,  \nonumber \\
L^{2}\frac{d}{dL}\left( L^{-1}I_{\varepsilon ^{2}}\right) &=&-\frac{15\phi }{%
4}-\frac{9\cos ^{4}\phi -26\cos ^{2}\phi +15}{4\sin ^{3}\phi }\cos \phi ~, 
\nonumber \\
L^{4}\frac{d}{dL}\left( L^{-3}I_{\eta ^{2}}\right) &=&-\frac{105\phi }{64} 
\nonumber \\
&&-\frac{6\cos ^{6}\phi +21\cos ^{4}\phi -105\cos ^{2}\phi +70}{64\sin
^{3}\phi }\cos \phi ~,  \nonumber \\
L^{3}\frac{d}{dL}\left( L^{-2}I_{\varepsilon \eta }\right) &=&\frac{8}{\sin
\phi }+\frac{\cos ^{4}\phi +6\cos ^{2}\phi -24}{2\sin \phi }\cos ^{2}\phi ~.
\end{eqnarray}%
As the last term of each expression contains the factor $\cos \phi $, only
the first terms will survive the limiting process described in subsection
3.1.$\ $\ The respective limits are given in subsection 3.2.


\section*{References}

\begin{thebibliography}{99}
\bibitem{RS} L Randall and R Sundrum 1999 \textit{Phys. Rev. Lett.} \textbf{%
83} 4690

\bibitem{BDEL} P Bin\'{e}truy, C Deffayet, U Ellwanger and D Langlois 2000 
\textit{Phys. Lett.} B \textbf{477} 285

\bibitem{MakHarko} M K Mak and T Harko 2004 \textit{Phys. Rev. D} \textbf{70}
024010

\bibitem{Kar} S Pal, S Bharadwaj and S Kar 2005 \textit{Phys. Lett. B} 
\textbf{609} 194

\bibitem{Pal} S Pal 2005 \textit{Phys. Teacher} \textbf{47} 144

\bibitem{SMS} T Shiromizu, K Maeda and M Sasaki 2000 \textit{Phys. Rev.} D 
\textbf{62} 024012

\bibitem{Decomp} L \'{A} Gergely 2003 \textit{Phys. Rev. D} \textbf{68}
124011

\bibitem{VarBraneTensionPRD} L \'{A} Gergely 2008 \textit{Phys. Rev.} D 
\textbf{78} 084006

\bibitem{BGNS} P Bostock, R Gregory, I Navarro and J Santiago 2004 \textit{%
Phys. Rev. Lett.} \textbf{92} 221601

\bibitem{deRham} C de Rham and A J Tolley 2006 \textit{JCAP} \textbf{0602}
003

\bibitem{Kaloper} N Kaloper and D Kiley 2006 \textit{JHEP} \textbf{0603} 077

\bibitem{KantiTamvakis} P Kanti and K Tamvakis 2002 \textit{Phys. Rev. D} 
\textbf{65} 084010

\bibitem{KantiOlasagastiTamvakis} P Kanti, I Olasagasti and K Tamvakis 2003 
\textit{Phys. Rev. D} \textbf{68} 124001

\bibitem{EmparanFabbriKaloper} R Emparan, A Fabbri and N Kaloper 2002 
\textit{JHEP} \textbf{0208} 043

\bibitem{KudohTanakaNakamura} H Kudoh, T Tanaka and T Nakamura 2003 \textit{%
Phys. Rev. D} \textbf{68} 024035

\bibitem{tidalRN} N Dadhich, R Maartens, P Papadopoulos and V Rezania 2000 
\textit{Phys. Lett. B} \textbf{487} 1

\bibitem{ChRH} A Chamblin, S W Hawking and H S Reall 2000 \textit{Phys. Rev.}
D \textbf{61} 065007

\bibitem{SCMlet} S S Seahra, C Clarkson and R Maartens 2005 \textit{Phys.
Rev. Lett} \textbf{94} 121302

\bibitem{GL} R Gregory and R Laflamme 1993 \textit{Phys. Rev. Lett}. \textbf{%
70} 2837

\bibitem{Gregory} R. Gregory, \textit{Class. Quantum Grav. }2000 \textbf{17}
L125

\bibitem{HorowitzMaeda} G T Horowitz and K Maeda 2001 \textit{Phys. Rev. Lett%
} \textbf{87} 131301

\bibitem{GarrigaTanaka} J Garriga and T Tanaka 2000 \textit{Phys. Rev. Lett.}
\textbf{84} 2778

\bibitem{Giddings} S B Giddings, E Katz and L Randall 2000 \textit{JHEP} 
\textbf{03} 023

\bibitem{BlackString} L \'{A} Gergely 2006 \textit{Phys. Rev. D} \textbf{74}
024002

\bibitem{GermaniMaartens} C Germani and R Maartens R 2001 \textit{Phys. Rev.
D} \textbf{64} 124010

\bibitem{BoehmerHarkoLobo} C G B\"{o}hmer, T Harko and F S N Lobo 2008 
\textit{Class. Quantum Grav}. \textbf{25} 045015

\bibitem{accretion} C S J Pun, Z Kovacs and T Harko 2008 \textit{Phys. Rev. D%
} \textbf{78} 084015

\bibitem{tidalThermo} L \'{A} Gergely, N Pidokrajt and S Winitzki 2008 
\textit{Thermodynamics of tidal charged black holes,} e-print:
arXiv:0811.1548

\bibitem{KarSinha} S Kar and M Sinha 2003 \textit{Gen. Rel. Grav.} \textbf{35%
} 10

\bibitem{MM} A S Majumdar and N Mukherjee 2005 \textit{Mod. Phys. Lett.} 
\textbf{A20} 2487

\bibitem{Whisker} R\ Whisker 2005 \textit{Phys. Rev. D} \textbf{71} 064004

\bibitem{MajumdarMukherjee} A S Majumdar and N Mukherjee 2005 \textit{Int.
J. Mod. Phys D} \textbf{14} 1095

\bibitem{SchaferBartelmann} B M Sch\"{a}fer and M Bartelmann 2006 \textit{%
Mon. Not. Roy. Astron. Soc.} \textbf{369} 425

\bibitem{GeDa} L \'{A} Gergely, B Dar\'{a}zs 2006 \textit{Publ. Astron.
Dept. E\"{o}tv\"{o}s Univ. PADEU} \textbf{17} 213 e-print:
arXiv:astro-ph/0602427

\bibitem{Hobill} J Bri\"{e}t and D Hobill 2008 \textit{Determining the
dimensionality of space-time by gravitational lensing}, e-print:
arXiv:0801.3859

\bibitem{Straumann} N Straumann 2004 General Relativity (Springer), ch. 3

\bibitem{Landau} L D Landau and E M Lifshitz 1979 \textit{The Classical
Theory of Fields} (Oxford: Butterworth-Heinemann)

\bibitem{RadioInt} D S Robertson , W E Carter and W H Dillinger 1991 \textit{%
Nature} \textbf{349} 768

\bibitem{RadioInt2} D E Lebach, B E Corey, I I Shapiro, M I Ratner, J C
Webber, A E E Rogers, J L Davis and T A Herring 1995 \textit{Phys. Rev. Lett.%
} \textbf{75} 1439

\bibitem{braneworldgrav} R Maartens 2004 \textit{Living Rev. Rel.} \textbf{7}
1

\bibitem{IrradFriedmann} L \'{A} Gergely and Z Keresztes 2006 \textit{JCAP} 
\textbf{01} 022

\bibitem{BBNLim} R Maartens, D Wands, B A Bassett and I P C Heard 2000 
\textit{Phys. Rev. D} \textbf{62} 041301(R)

\bibitem{Sereno} M Sereno 2003 \textit{Phys. Rev. D} \textbf{67} 064007
\end{thebibliography}
\end{document}